\documentclass[11pt]{article}

\usepackage[preprint]{acl}

\usepackage{times}
\usepackage{latexsym}

\usepackage[T1]{fontenc}

\usepackage[utf8]{inputenc}

\usepackage{microtype}
\usepackage{inconsolata}

\usepackage{graphicx}
\usepackage{amsmath,amssymb}
\usepackage{booktabs}
\usepackage{tabularx}
\usepackage{multirow}
\usepackage{xcolor}
\usepackage{colortbl}
\usepackage{stfloats}

\definecolor{sjHeaderGray}{HTML}{F2F2F2}
\definecolor{sjOursBlue}{HTML}{EAF2FB}

\newcommand{\sjNA}{\textemdash}
\newcommand{\sjBest}[1]{\textbf{#1}}
\newcommand{\sjSecond}[1]{\underline{#1}}

\newcommand{\sjScores}[4]{#1 & #2 & #3 & #4}

\newlength{\sjValueWidth}
\newlength{\sjFprWidth}

\newcommand{\sjMetricHead}[3][\sjValueWidth]{%
    \shortstack{#2\\\makebox[#1][c]{#3}}%
}

\newcommand{\sjMetricHeads}{%
    \sjMetricHead{AUC}{$\uparrow$} &
    \sjMetricHead{OSCR}{$\uparrow$} &
    \sjMetricHead{Acc}{$\uparrow$} &
    \sjMetricHead[\sjFprWidth]{FPR}{$\downarrow$}%
}

\title{Teaching Pretrained Language Models to Abstain on Unseen Radio Signals via Complementary Candidate Lists}
\author{
  Ruixiang Zhang \quad Zinan Zhou \quad Yezhuo Zhang \quad Xuanpeng Li\thanks{Corresponding author.} \\
  \normalfont\normalsize School of Instrument Science and Engineering, Southeast University, Nanjing 210096, China \\
  \normalfont\normalsize State Key Laboratory of Comprehensive PNT Network and Equipment Technology \\
  \normalfont\normalsize Key Laboratory of Micro-Inertial Instrument and Advanced Navigation Technology, Ministry of Education \\
  \normalfont\small\texttt{\{zhang\_ruixiang, zhouzinan919, zhang\_yezhuo, li\_xuanpeng\}@seu.edu.cn}
}

\begin{document}
\maketitle

\begin{abstract}
Open-set radio signal recognition requires classifying known modulation schemes while abstaining on unseen types. One of the main challenges lies in a fundamental supervision gap: when training, model selection, and calibration access only known classes, no example directly supervises rejection of unseen categories, leaving closed-set classifiers prone to overconfident errors and split-dependent collapses. To address this, we propose OSCAL\footnote{Acronym for \textbf{O}pen-\textbf{S}et \textbf{C}andidate-conditioned \textbf{A}bstention \textbf{L}earning.}, an abstention learning framework that adapts a frozen pretrained language model (LM) to continuous waveforms via complementary candidate-conditioned instruction tuning (OpenSFT). Building on dual-view signal tokens initialized by masked reconstruction, OpenSFT pairs each known waveform with disjoint candidate subsets to supervise class selection when listed and \texttt{UNKNOWN} when omitted within the native vocabulary space. We further introduce native-likelihood margin scoring with known-only predicted-class standardization to align margin scales across modulation families. Extensive experiments across four open-set protocols and 18 baselines demonstrate strong cross-protocol robustness without split-specific collapses. Further analysis highlights that language pretraining converts partial-list abstention from a shortcut into unseen-class transfer, and characterizes how signal reconstruction, score standardization, and modulation proximity govern rejection.
\end{abstract}

\section{Introduction}
\begin{figure}[t]
    \centering
    \includegraphics[width=0.90\columnwidth]{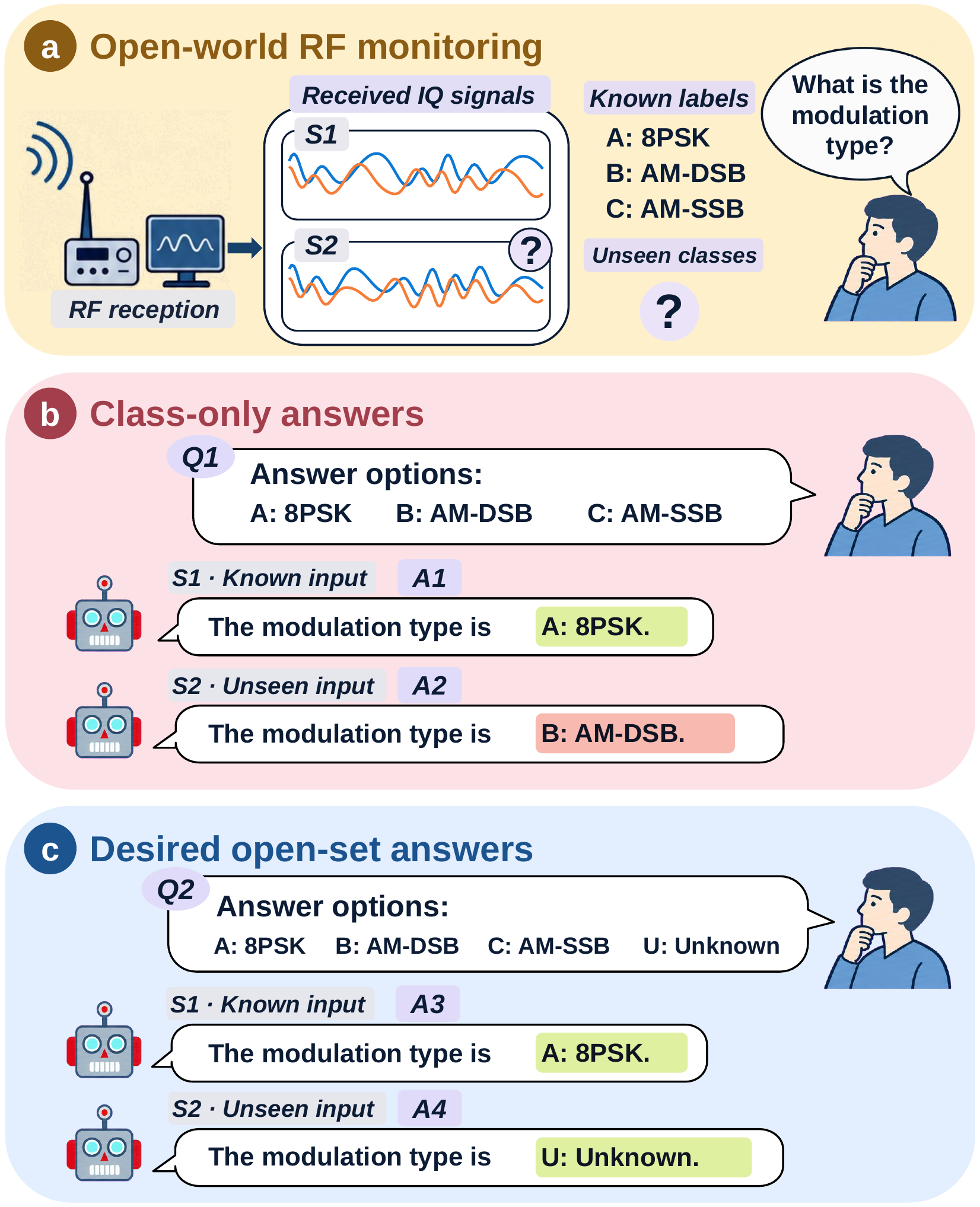}
    \caption{Closed-set versus open-set RF recognition: (a)~open-world monitoring encounters both known signal S1 and unseen signal S2; (b)~class-only options force S2 into a false known label; (c)~including \texttt{U: Unknown} enables known classification on S1 and abstention on S2.}
    \label{fig:motivation}
\end{figure}

Automatic modulation classification (AMC) identifies how information is encoded in a received radio-frequency (RF) waveform. In open-world spectrum monitoring, receivers also encounter unseen modulation schemes absent from training, requiring recognizers to label familiar signals while abstaining via \texttt{UNKNOWN} on unfamiliar ones (Figure~\ref{fig:motivation}). Yet when training, selection, and calibration access only known classes $\mathcal{K}$, a fundamental supervision gap arises: no unseen-class waveform exists to teach abstention. Because continuous RF waveforms lie outside text pretraining corpora, this setting also offers a contamination-free testbed for studying how pretrained language models (LMs) transfer abstention to genuinely unseen categories.

Conventional open-set recognizers---including post-hoc confidence rules, feature-space OSR modules, and detached-head RF--LLMs such as RadioLLM \citep{radiollm}---rely on fixed-$K$ classification heads or per-class feature modules and cannot dynamically condition inference on arbitrary candidate subsets, whereas text-prompting RF--LLMs \citep{context_is_king,disc_amc} replace continuous waveform tokens with verbalized or discretized features. Adapting a pretrained LM with continuous RF tokens instead enables candidate-conditioned answering: pairing any subset with \texttt{UNKNOWN} in the native vocabulary space supervises each known signal to select its class when listed and abstain when omitted. Yet partial-list abstention does not guarantee unseen-class rejection. A model can fit subset supervision via a shortcut---classifying an input into its closest known class and outputting \texttt{UNKNOWN} only when that class is omitted---which still misclassifies unseen signals when all known classes are listed at deployment. This raises two questions: \textbf{(RQ1)} Can complementary candidate-conditioned \texttt{UNKNOWN} supervision on known waveforms transfer to unseen-class rejection? \textbf{(RQ2)} How do language pretraining, masked reconstruction, and the native answer interface govern this transfer?

In this work, we propose \textbf{OSCAL} to bridge candidate-omission supervision and open-set rejection. OSCAL tokenizes waveforms into temporal IQ and spectral PSD views, initializing the RF--LM pathway via masked reconstruction (\S\ref{sec:stage1}). \textbf{OpenSFT} then pairs each known signal with disjoint half-subsets of $\mathcal{K}$ to supervise class codes and \texttt{UNKNOWN} through the frozen tied vocabulary head (\S\ref{sec:opensft}), while inference applies known-only predicted-class $z$-standardization to native-likelihood margins (\S\ref{sec:scoring}).

Across four nested RML22 open-set protocols (u2--u5) and 18 baselines, OSCAL achieves the strongest cross-protocol robustness, lifting mean OSCR by 6.67 and worst-case OSCR by 7.86 points over the best comparator without split-specific collapses. Controlled ablations confirm that language pretraining unlocks unseen-class transfer specifically through OpenSFT's candidate interface, and diagnostic analyses (\S\ref{sec:analysis}) characterize the calibration, physical, and contextual boundaries of rejection.

\section{Related Work}
\paragraph{RF Adaptation of Language Models.}
Prior RF--LLM research follows either continuous alignment or text-only prompting. Continuous-alignment approaches map waveforms or electromagnetic views into LLM hidden spaces via prompt/token reprogramming with an external classifier head \citep{radiollm}, prototype-guided Q-Former alignment \citep{radiollm_v2}, dual-domain signal encoding with knowledge-enhanced instructions \citep{amr_llm}, or multi-view visual encodings \citep{pred}. Text-only pipelines instead verbalize statistics or retrieve discretized exemplars into frozen or QLoRA-tuned LLMs \citep{context_is_king,disc_amc} rather than encoding continuous waveforms directly. Unlike detached-head or text-only systems, OSCAL pairs continuous dual-view RF tokens with candidate-conditioned prompts and supervises known classes and \texttt{UNKNOWN} directly in the frozen native vocabulary space.

\paragraph{Open-Set Recognition and Modulation Classification.}
Open-set recognition encompasses post-hoc logit scoring---using probability, logits, energy, or extreme-value recalibration \citep{bendale2016openmax,hendrycks2017msp,liu2020energy,vaze2022mls}---alongside representation-shaping methods built on class-specific reconstruction, ensembles, or generative boundaries \citep{huang2023cssr,wang2024medaf,liu2024csgrl}. RF-specific open-set networks further enforce class-space reconstruction \citep{chen2026cfsr}, local--global feature-space center constraints \citep{zhang2026lgfscn}, or manifold-mixup placeholders \citep{zhang2026mlfdl}. Operating on fixed-$K$ feature-space or logit geometry without a candidate-conditioned abstention interface, these approaches still fluctuate sharply across unseen partitions.

\paragraph{Abstention in Language Models.}
NLP abstention spans unanswerable and selective QA \citep{rajpurkar2018squad2,kamath2020selective}, MCQA analyses exposing option-symbol selection biases \citep{zheng2024mcqa} and ``None of the Above'' failure modes \citep{tam2025nota}, token-level rejection tuning (SEAL; \citealp{huang2025seal}), and electromagnetic MCQA with an ``Unable to answer'' option for ambiguous inputs \citep{pred}. Rather than relying on token-level mispredictions or curated refusal examples, OSCAL derives \texttt{UNKNOWN} supervision deterministically from candidate omission on known waveforms and tests transfer to genuinely unseen classes.

\section{Method}
\subsection{Problem Formulation and Overview}
\begin{figure*}[t]
    \centering
    \includegraphics[width=0.98\textwidth]{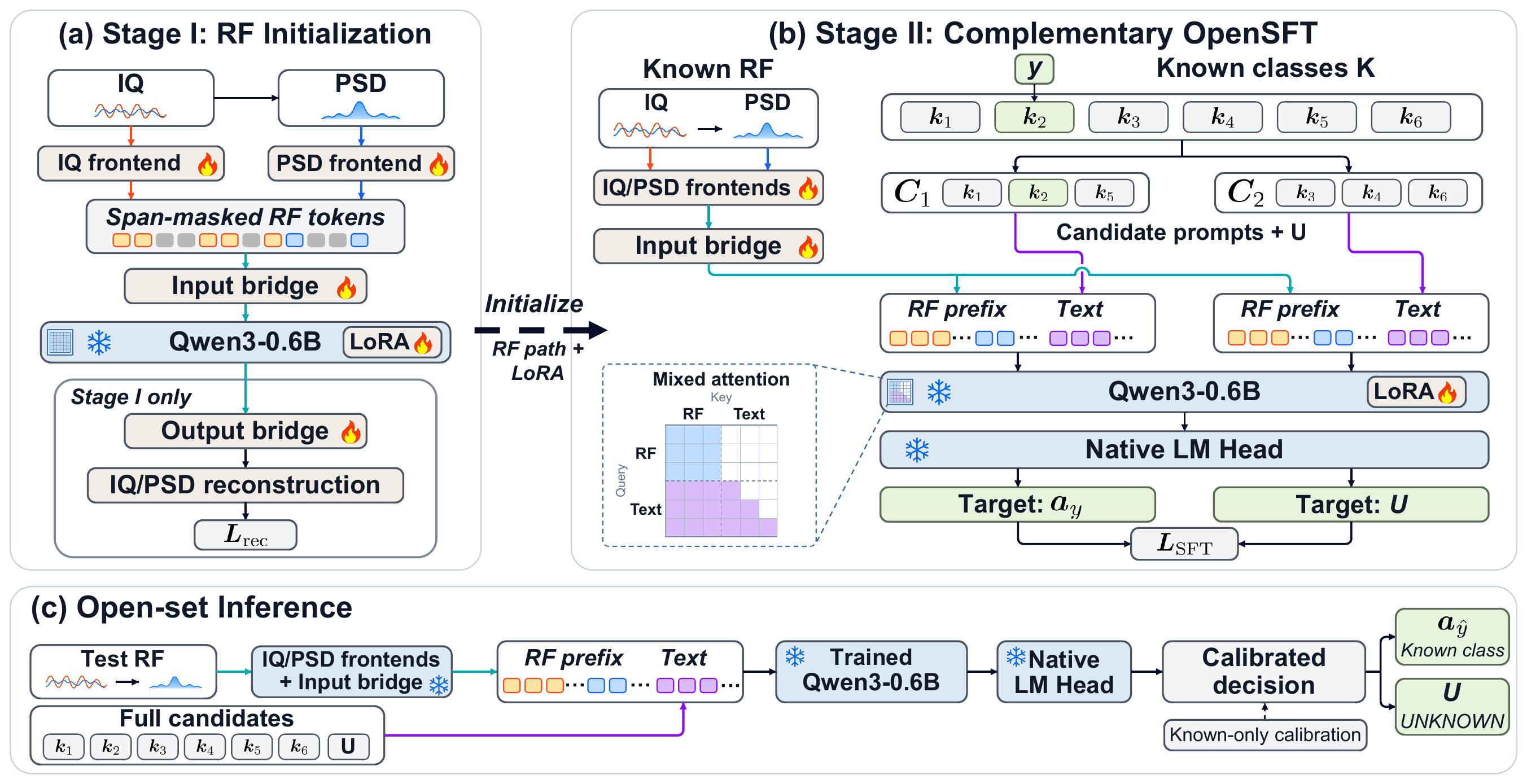}
    \caption{Overview of OSCAL: (a)~Stage~I (\S\ref{sec:stage1}) aligns dual-view RF tokens and LoRA via masked IQ/PSD reconstruction under bidirectional attention; (b)~Stage~II OpenSFT (\S\ref{sec:opensft}) pairs each known waveform with disjoint half-subsets of $\mathcal{K}$ under mixed attention, supervising $a_y$ or \texttt{U} through the frozen native LM head; (c)~open-set inference (\S\ref{sec:scoring}) evaluates full candidates with known-only calibration to output $a_{\hat{y}}$ or \texttt{UNKNOWN}. Flames and snowflakes mark trainable and frozen modules.}
    \label{fig:overview}
\end{figure*}

Let $x \in \mathbb{R}^{L \times 2}$ denote an in-phase/quadrature (IQ) waveform, $\mathcal{K}$ the known modulation classes, and $\mathcal{U}$ disjoint unseen test-only classes ($\mathcal{K} \cap \mathcal{U} = \emptyset$). Training, checkpoint selection (Selection-Val), and calibration (Known-Cal) access only known samples ($y \in \mathcal{K}$). Given a prompt $P(\mathcal{C})$ listing a candidate subset $\mathcal{C} \subseteq \mathcal{K}$ alongside \texttt{UNKNOWN}, the target is $y$'s class code if $y \in \mathcal{C}$ and \texttt{UNKNOWN} otherwise. Crucially, OpenSFT trains on known waveforms under partial lists ($\mathcal{C} \subset \mathcal{K}$), whereas inference evaluates mixed known and unseen waveforms against the full known set ($\mathcal{C} = \mathcal{K}$). Figure~\ref{fig:overview} depicts the full pipeline, and Table~\ref{tab:prompt_example} illustrates the prompt--target format on u4 (verbatim template and all 11 modulation descriptions in Appendix~\ref{app:prompts}).

\begin{table}[t]
\centering
\scriptsize
\setlength{\tabcolsep}{3.5pt}
\renewcommand{\arraystretch}{1.10}
\begin{tabularx}{\columnwidth}{X}
\toprule
\rowcolor{sjHeaderGray}
\textbf{Shared System Instruction} \\
\midrule
\texttt{You are an expert in radio modulation classification. Analyze the input RF signal and select its modulation type from the given candidate categories.} \\
\midrule
\rowcolor{sjHeaderGray}
\textbf{Stage~II Complementary Training Pair} (known $y=\text{BPSK}$, u4) \\
\midrule
\textbf{Subset $\mathcal{C}_1$ (contains $y$):} \texttt{Candidate categories:} \\
\texttt{B. AM-DSB: analog amplitude modulation...} \\
\texttt{D. BPSK: digital PSK; 2 carrier phase states; constant envelope...} \\
\texttt{F. GFSK: digital FSK...} \quad \texttt{U. UNKNOWN: None of the listed modulation types.} \\
\texttt{Which category does the RF signal belong to? Answer:} $\rightarrow$ \textbf{Target: \texttt{D}} \\
\addlinespace[1.5pt]
\textbf{Subset $\mathcal{C}_2$ (omits $y$):} \texttt{Candidate categories:} \\
\texttt{C. AM-SSB: ...} \; \texttt{E. CPFSK: ...} \; \texttt{G. PAM4: ...} \; \texttt{K. WBFM: ...} \\
\texttt{U. UNKNOWN: None of the listed modulation types.} \\
\texttt{Which category does the RF signal belong to? Answer:} $\rightarrow$ \textbf{Target: \texttt{U}} \\
\midrule
\rowcolor{sjOursBlue}
\textbf{Open-Set Inference Prompt $P(\mathcal{K})$} \\
\midrule
Lists \texttt{B, C, D, E, F, G, K} + \texttt{U. UNKNOWN}. Compute $m(x) = \ell_\theta(a_{\hat{y}}) - \ell_\theta(a_U)$; accept $\hat{y}$ if $z(x) \ge \tau$, else \texttt{UNKNOWN}. \\
\bottomrule
\end{tabularx}
\caption{Prompt--target structure in OSCAL on u4 for known BPSK. Single-token codes (\texttt{A}--\texttt{K}, \texttt{U}) remain invariant across protocols; descriptions are abbreviated (verbatim text in Appendix~\ref{app:prompts}).}
\label{tab:prompt_example}
\end{table}

\subsection{Stage~I: RF Initialization via Masked Reconstruction}
\label{sec:stage1}
Modulation schemes differ in both temporal constellations and spectral envelopes. To capture both cues, two lightweight 1D convolutional frontends tokenize each RMS-normalized IQ waveform into 32 temporal IQ tokens and 8 STFT log power spectral density (PSD) tokens, forming a 40-token RF prefix with view, positional, and projected time--frequency coordinate embeddings.

To regularize known-signal representations prior to instruction tuning, Stage~I initializes the RF--LM pathway on frozen \texttt{Qwen3-0.6B-Base} \citep{yang2025qwen3} via label-free reconstruction. Contiguous IQ spans and a spectral PSD band are masked with learned mask embeddings; a linear input bridge projects the 40 tokens into the LM under full bidirectional attention while LoRA adapters \citep{lora} update attention and feed-forward projections. A Stage~I-only output bridge and view heads predict per-sample standardized masked patches $t_i^v \in \mathbb{R}^{d_v}$ over masked index sets $M_v$:
\begin{align}
\mathcal{L}_v &= \frac{1}{|M_v|\, d_v} \sum_{i \in M_v} \|\hat{t}_i^v - t_i^v\|_2^2, \\
\mathcal{L}_{\mathrm{rec}} &= \mathcal{L}_{\mathrm{IQ}} + 0.25\,\mathcal{L}_{\mathrm{PSD}}.
\end{align}
See Appendix~\ref{app:impl} for architectural and masking details.

\subsection{Stage~II: Complementary OpenSFT}
\label{sec:opensft}
To supervise abstention from known signals alone, OpenSFT pairs each known sample $(x, y)$ at every epoch with two disjoint nonempty half-subsets $\mathcal{C}_1, \mathcal{C}_2$ of $\mathcal{K}$ of sizes $\lfloor |\mathcal{K}|/2 \rfloor$ and $\lceil |\mathcal{K}|/2 \rceil$, ensuring $y$ appears in exactly one subset. As illustrated in Table~\ref{tab:prompt_example}, each prompt $P(\mathcal{C}_q)$ formats active candidates as structured code--name--description entries alongside \texttt{U. UNKNOWN} using protocol-invariant single-token codes \texttt{A}--\texttt{K} and \texttt{U}, avoiding multi-token length bias and re-indexing. Section~\ref{sec:discussion} further examines how predictions rely on these symbolic anchors versus modulation descriptions. Training partitions vary across samples and epochs to prevent subset memorization, while Selection-Val partitions stay fixed.

Rather than attaching a new classification head, OSCAL scores each legal single-token answer $a$ through the frozen tied projection over the \emph{full} LM vocabulary:
\begin{equation}
\ell_\theta(a; x, \mathcal{C}) = \log p_\theta\big(a \mid x, P(\mathcal{C})\big),
\end{equation}
minimizing the balanced negative log-likelihood over complementary targets $a_1, a_2$ (one known class code, one \texttt{U}):
\begin{equation}
\mathcal{L}_{\mathrm{SFT}} = -\frac{1}{2}\sum_{q=1}^{2} \ell_\theta(a_q; x, \mathcal{C}_q).
\end{equation}
Retaining the frozen tied vocabulary head preserves pretrained choice geometry while updating only the RF frontend, input bridge, and LoRA weights. A mixed attention mask applies bidirectional attention within the 40 RF tokens and causal attention from text tokens to the RF prefix and prior text, preventing prompts from altering RF encodings.

\subsection{Native-Likelihood Scoring and Calibration}
\label{sec:scoring}
At inference, OSCAL evaluates the full known prompt $P(\mathcal{K})$ with \texttt{UNKNOWN} to compute the predicted class and raw acceptance margin:
\begin{align}
\hat{y}(x) &= \arg\max_{k \in \mathcal{K}} \ell_\theta(a_k; x, \mathcal{K}), \\
m(x) &= \ell_\theta(a_{\hat{y}}; x, \mathcal{K}) - \ell_\theta(a_U; x, \mathcal{K}),
\end{align}
which equals the native logit difference $s_{\hat{y}}(x) - s_U(x)$ because all single-token answers share one log-partition function at the answer boundary.

However, raw margin scales vary widely across modulation families, so a single global threshold on $m(x)$ conflates low-margin known classes with unseen signals. Fitting strictly on Known-Cal, we group samples by predicted class $\hat{y}(x)$ to match inference routing and standardize margins via bucket statistics $(\mu_k, \sigma_k)$:
\begin{equation}
\label{eq:class_z}
z(x) = \frac{m(x) - \mu_{\hat{y}(x)}}{\sigma_{\hat{y}(x)}},
\end{equation}
with global Known-Cal fallback for sparse buckets (Appendix~\ref{app:impl}). Setting $\tau$ to the 5th percentile of Known-Cal $z(x)$ accepts $\hat{y}(x)$ when $z(x) \ge \tau$ and outputs \texttt{UNKNOWN} otherwise.

\section{Experimental Setup}
\subsection{Dataset and Open-Set Protocols}
We evaluate on the 0--20~dB subset (in 2~dB steps) of RML22 \citep{sathyanarayanan2023rml22} (11 modulation classes, 128-sample complex IQ waveforms) under our fixed stratified 60\%/20\%/20\% split. Unseen classes are strictly excluded from all non-test folds, and the known validation fold is split equally by class and SNR into Selection-Val and Known-Cal to ensure zero unseen leakage. We construct four nested open-set protocols denoted u$m$ ($m \in \{2,3,4,5\}$), each holding out $m$ of the 11 classes as unseen to progressively stress-test physical proximity: u2 ($\mathcal{U}=\{\text{QAM16}, \text{QAM64}\}$) holds out high-order QAM; u3 adds 8PSK while keeping adjacent QPSK known; u4 removes QPSK as well; and u5 additionally holds out analog AM-DSB alongside known WBFM and AM-SSB (split counts in Appendix~\ref{app:prompts}).

\subsection{Baselines}
Table~\ref{tab:openset_main} compares 18 baselines across four families on identical splits (full recipes in Appendix~\ref{app:baselines}):

\paragraph{Prompted-CE post-hoc scores (Group~a).}
MSP \citep{hendrycks2017msp}, MLS \citep{vaze2022mls}, Energy \citep{liu2020energy}, and OpenMax \citep{bendale2016openmax} on the identical Stage~I \texttt{Qwen3-0.6B-Base} checkpoint fine-tuned with closed-set CE over $P(\mathcal{K})$ (without \texttt{UNKNOWN}), isolating OpenSFT's gain over post-hoc scoring.

\paragraph{Qwen-adapted OSR modules (Group~b).}
CSSR \citep{huang2023cssr}, MEDAF \citep{wang2024medaf}, and CSGRL \citep{liu2024csgrl} attached to the shared RF-Qwen backbone, contrasting detached feature-space OSR modules with the native vocabulary interface.

\paragraph{AMC-specific OSR networks (Group~c).}
Domain-specialized classifiers: CFSR \citep{chen2026cfsr} (ResNet18-1D; \citealp{he2016resnet}), LGFSCN \citep{zhang2026lgfscn} (LGFE), and MLFDL \citep{zhang2026mlfdl} (MCLDNN; \citealp{xu2020mcldnn}).

\paragraph{LLM-based AMC pipelines (Group~d).}
Text-prompting systems---Context Is King \citep{context_is_king} and DiSC-AMC \citep{disc_amc} (\texttt{TF, Centroid}, \texttt{TF, RAG}, \texttt{QLoRA, RAG})---on \texttt{Qwen3-0.6B-Base} and \texttt{DeepSeek-R1-Distill-Qwen-7B} (\texttt{DS-R1-7B}; \citealp{deepseek2025r1}).

Each baseline retains its official open-set scoring function, with any auxiliary score statistics fitted strictly on known non-test folds.

\subsection{Evaluation Metrics and Implementation}
We report four metrics on a 0--100 scale: pre-rejection known accuracy (\textbf{Acc}); threshold-free Area Under the ROC Curve (\textbf{AUROC}, shown as \textbf{AUC}) for known-versus-unseen discrimination; Open-Set Classification Rate (\textbf{OSCR}), jointly measuring known classification and unseen rejection across thresholds; and False Positive Rate at 95\% known True Positive Rate (\textbf{FPR95}, shown as \textbf{FPR}) at the 95\%-TPR ROC operating point. OSCAL updates 10.4M LoRA and RF parameters (1.7\% of 606.7M) in BF16 with FP32 vocabulary scoring, selecting checkpoints by minimum Selection-Val $\mathcal{L}_{\mathrm{SFT}}$; full optimization schedules appear in Appendix~\ref{app:impl}.

\section{Results}

\subsection{Main Results}
\begin{table*}[t]
\centering

\fontsize{8}{9.5}\selectfont
\setlength{\tabcolsep}{1.2pt}
\renewcommand{\arraystretch}{1.16}

\settowidth{\sjValueWidth}{00.00}
\settowidth{\sjFprWidth}{100.00}

\resizebox{\textwidth}{!}{%
\begin{tabular}{
    l
    >{\hspace{2pt}}l
    *{16}{c}
}
\toprule

\rowcolor{sjHeaderGray}
& & \multicolumn{4}{c}{u2}
  & \multicolumn{4}{c}{u3}
  & \multicolumn{4}{c}{u4}
  & \multicolumn{4}{c}{u5} \\

\cmidrule(lr){3-6}
\cmidrule(lr){7-10}
\cmidrule(lr){11-14}
\cmidrule(lr){15-18}

\rowcolor{sjHeaderGray}
Method & Backbone
& \sjMetricHeads
& \sjMetricHeads
& \sjMetricHeads
& \sjMetricHeads \\

\midrule

\multicolumn{18}{l}{
    \textit{(a) Prompted-CE post-hoc scores}
} \\[1pt]

MSP & Qwen3-0.6B
& \sjScores{89.50}{88.20}{\sjSecond{96.98}}{51.73} 
& \sjScores{65.04}{64.81}{\sjSecond{98.02}}{91.80} 
& \sjScores{67.01}{66.65}{\sjSecond{97.92}}{81.36} 
& \sjScores{71.95}{71.77}{\sjSecond{99.22}}{55.94} \\ 

MLS & Qwen3-0.6B
& \sjScores{\sjSecond{90.58}}{\sjBest{89.12}}{\sjSecond{96.98}}{\sjBest{47.73}} 
& \sjScores{\sjBest{80.36}}{\sjBest{79.96}}{\sjSecond{98.02}}{90.67} 
& \sjScores{59.56}{59.09}{\sjSecond{97.92}}{74.32} 
& \sjScores{74.31}{74.12}{\sjSecond{99.22}}{68.41} \\ 

Energy & Qwen3-0.6B
& \sjScores{\sjBest{90.59}}{\sjBest{89.12}}{\sjSecond{96.98}}{\sjBest{47.73}} 
& \sjScores{\sjBest{80.36}}{\sjBest{79.96}}{\sjSecond{98.02}}{90.72} 
& \sjScores{59.53}{59.05}{\sjSecond{97.92}}{74.03} 
& \sjScores{74.30}{74.11}{\sjSecond{99.22}}{68.45} \\ 

OpenMax & Qwen3-0.6B
& \sjScores{89.34}{88.05}{96.68}{\sjSecond{50.28}} 
& \sjScores{64.73}{64.49}{97.91}{\sjSecond{89.49}} 
& \sjScores{63.81}{63.41}{97.91}{78.62} 
& \sjScores{68.38}{68.18}{98.94}{60.52} \\ 

\addlinespace[2.5pt]

\multicolumn{18}{l}{
    \textit{(b) Qwen-adapted OSR methods}
} \\[1pt]

Qwen-CSSR & Qwen3-0.6B
& \sjScores{47.24}{46.54}{96.73}{92.27} 
& \sjScores{44.04}{41.42}{96.11}{99.70} 
& \sjScores{82.20}{79.79}{97.46}{84.69} 
& \sjScores{52.22}{51.70}{99.16}{88.14} \\ 

Qwen-MEDAF & Qwen3-0.6B
& \sjScores{83.36}{82.43}{\sjBest{97.00}}{76.39} 
& \sjScores{51.82}{51.72}{\sjBest{98.25}}{96.02} 
& \sjScores{78.59}{78.01}{\sjBest{98.22}}{66.01} 
& \sjScores{70.81}{70.68}{\sjBest{99.34}}{56.65} \\ 

Qwen-CSGRL & Qwen3-0.6B
& \sjScores{66.49}{64.98}{96.53}{96.03} 
& \sjScores{37.68}{37.27}{97.72}{95.92} 
& \sjScores{48.03}{46.79}{97.37}{87.90} 
& \sjScores{67.91}{67.49}{99.16}{83.96} \\ 

\addlinespace[2.5pt]

\multicolumn{18}{l}{
    \textit{(c) AMC-specific methods}
} \\[1pt]

CFSR & ResNet18-1D
& \sjScores{58.97}{57.90}{94.18}{73.90} 
& \sjScores{71.28}{70.47}{96.79}{91.74} 
& \sjScores{71.63}{70.07}{96.66}{55.53} 
& \sjScores{79.98}{79.18}{97.02}{\sjBest{30.84}} \\ 

LGFSCN & LGFE
& \sjScores{77.65}{76.39}{96.30}{89.35} 
& \sjScores{66.97}{65.63}{96.47}{91.37} 
& \sjScores{\sjBest{92.53}}{\sjSecond{90.10}}{96.35}{51.00} 
& \sjScores{68.32}{68.15}{98.94}{85.42} \\ 

MLFDL & MCLDNN
& \sjScores{79.32}{78.12}{96.92}{88.16} 
& \sjScores{60.58}{59.87}{97.47}{91.24} 
& \sjScores{89.16}{87.37}{97.48}{\sjSecond{41.76}} 
& \sjScores{\sjSecond{81.87}}{\sjBest{81.53}}{99.07}{\sjSecond{54.11}} \\ 

\addlinespace[2.5pt]

\multicolumn{18}{l}{
    \textit{(d) LLM-based AMC methods}
} \\[1pt]

\multirow{2}{*}{Context Is King}
& Qwen3-0.6B
& \sjScores{52.57}{5.90}{11.04}{94.18} 
& \sjScores{57.17}{6.27}{11.82}{93.43} 
& \sjScores{53.58}{7.45}{12.57}{93.56} 
& \sjScores{47.35}{6.81}{15.10}{95.74} \\ 

& DS-R1-7B
& \sjScores{44.85}{3.61}{8.63}{94.94} 
& \sjScores{40.48}{3.39}{10.73}{99.01} 
& \sjScores{42.64}{4.54}{13.17}{97.20} 
& \sjScores{53.07}{6.56}{16.39}{92.11} \\ 

\addlinespace[1.5pt]

\multirow{2}{*}{
    \shortstack[l]{DiSC-AMC\\(TF, Centroid)}
}
& Qwen3-0.6B
& \sjScores{61.21}{32.88}{47.00}{86.72} 
& \sjScores{58.62}{34.68}{55.62}{92.65} 
& \sjScores{53.46}{35.09}{58.38}{96.57} 
& \sjScores{60.99}{41.89}{61.89}{95.40} \\ 

& DS-R1-7B
& \sjScores{53.29}{14.91}{20.80}{98.92} 
& \sjScores{50.18}{17.74}{26.37}{98.39} 
& \sjScores{44.76}{21.72}{33.26}{99.04} 
& \sjScores{53.66}{25.97}{37.56}{91.34} \\ 

\addlinespace[1.5pt]

\multirow{2}{*}{
    \shortstack[l]{DiSC-AMC\\(TF, RAG)}
}
& Qwen3-0.6B
& \sjScores{62.12}{38.70}{56.99}{91.49} 
& \sjScores{53.54}{34.79}{57.04}{92.52} 
& \sjScores{49.49}{31.73}{55.90}{92.81} 
& \sjScores{54.35}{39.32}{66.66}{94.63} \\ 

& DS-R1-7B
& \sjScores{52.27}{17.56}{25.19}{92.55} 
& \sjScores{54.39}{20.82}{28.27}{95.71} 
& \sjScores{45.41}{21.18}{33.74}{98.20} 
& \sjScores{53.57}{24.26}{34.18}{94.19} \\ 

\addlinespace[1.5pt]

\multirow{2}{*}{
    \shortstack[l]{DiSC-AMC\\(QLoRA, RAG)}
}
& Qwen3-0.6B
& \sjScores{46.68}{11.38}{31.96}{99.36} 
& \sjScores{45.47}{9.81}{19.92}{96.08} 
& \sjScores{43.39}{10.04}{26.39}{95.24} 
& \sjScores{48.36}{1.40}{3.70}{96.27} \\ 

& DS-R1-7B
& \sjScores{56.12}{15.86}{27.20}{91.77} 
& \sjScores{45.67}{16.60}{35.27}{92.86} 
& \sjScores{56.28}{21.92}{33.83}{91.81} 
& \sjScores{45.89}{21.56}{42.29}{94.65} \\ 

\midrule

\rowcolor{sjOursBlue}
\textbf{OSCAL (ours)} & Qwen3-0.6B
& \sjScores{90.54}{\sjSecond{88.66}}{96.64}{55.51} 
& \sjScores{\sjSecond{74.27}}{\sjSecond{73.49}}{97.97}{\sjBest{70.89}} 
& \sjScores{\sjSecond{91.61}}{\sjBest{90.23}}{97.71}{\sjBest{41.51}} 
& \sjScores{\sjBest{81.95}}{\sjSecond{81.18}}{93.25}{89.97} \\ 

\bottomrule
\end{tabular}%
}

\caption{Open-set recognition results across four RML22 protocols (u2--u5), pooled over 0--20~dB. Bold and underlined numbers mark the best and second-best distinct scores per column; blue shading highlights OSCAL.}
\label{tab:openset_main}
\end{table*}

\begin{figure}[t]
    \centering
    \includegraphics[width=\columnwidth]{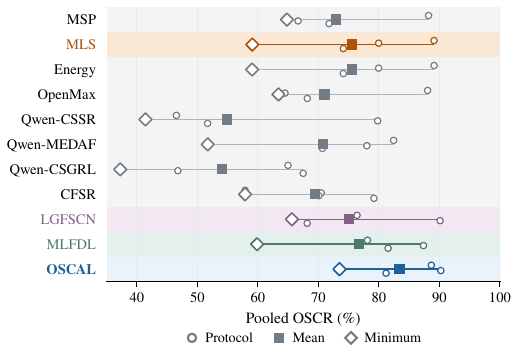}
    \caption{Pooled OSCR across u2--u5 (circles), mean (squares), and minimum (diamonds) for all 11 continuous-RF methods in Table~\ref{tab:openset_main}.}
    \label{fig:oscr_protocols}
\end{figure}

\paragraph{OSCAL achieves the strongest cross-protocol robustness.}
As shown in Table~\ref{tab:openset_main} and Figure~\ref{fig:oscr_protocols}, open-set performance fluctuates sharply across protocols for every baseline family: each comparator collapses on at least one partition, whereas OSCAL remains competitive across all four splits. Averaged over u2--u5, OSCAL attains the highest mean OSCR (83.39 vs.\ 76.72 for MLFDL) and lifts worst-case OSCR by 7.86 points (73.49 vs.\ 65.63 for LGFSCN), confirming that complementary candidate supervision mitigates split-specific collapses.

\paragraph{Baseline-family contrasts clarify the source of gains.}
On the identical Stage~I backbone, replacing closed-set Prompted-CE (Group~a) with OpenSFT eliminates MLS/Energy's severe u4 collapse (+31.14 OSCR over MLS). Although MLS leads global AUROC/OSCR on u3, it fails at 95\% TPR (90.67 FPR95), whereas OSCAL cuts u3 false acceptance by nearly 20 points (70.89 FPR95), confirming better-behaved high-recall margins. Qwen-adapted OSR modules (Group~b) retain high known accuracy yet falter on open-set separation, indicating that detached feature-space modules forfeit the native LM answer bias. Against specialized RF networks (Group~c), OSCAL matches or surpasses their u4/u5 peaks without their u2/u3 drops, while text-prompting pipelines (Group~d) fail across every split even at 7B or with QLoRA, confirming continuous RF tokenization is indispensable.

\begin{figure}[t]
    \centering
    \includegraphics[width=\columnwidth]{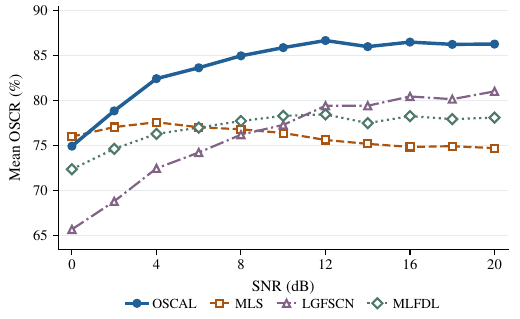}
    \caption{Mean OSCR across SNRs (averaged over u2--u5) for four representative methods.}
    \label{fig:oscr_snr}
\end{figure}

\paragraph{SNR stability and u5 decoupling.}
As shown in Figure~\ref{fig:oscr_snr}, while MLS holds a slight edge at 0~dB, OSCAL leads MLS, LGFSCN, and MLFDL across all SNRs from 2 to 20~dB and avoids MLS's high-SNR overconfidence drop above 4~dB. On u5, OSCAL still leads all methods in AUROC (81.95) despite lower Acc and higher FPR95, a ranking-versus-operating-point gap dissected in Section~\ref{sec:error_analysis}.

\subsection{Ablation Study}
\label{sec:ablation}
\begin{table}[t]
\centering\footnotesize
\setlength{\tabcolsep}{2.2pt}
\renewcommand{\arraystretch}{1.15}
\begin{tabularx}{\columnwidth}{ccc*{4}{>{\centering\arraybackslash}X}}
\toprule
\rowcolor{sjHeaderGray}
& & \multicolumn{1}{c}{Val} & \multicolumn{4}{c}{Open-Set Test} \\
\cmidrule(lr){3-3}\cmidrule(lr){4-7}
\rowcolor{sjHeaderGray}
\shortstack{Stage\\I} & \shortstack{LM\\Pretrained}
& \shortstack{Sub-U\\$\uparrow$}
& \shortstack{AUC\\$\uparrow$} & \shortstack{OSCR\\$\uparrow$}
& \shortstack{Acc\\$\uparrow$} & \shortstack{FPR\\$\downarrow$} \\
\midrule
\multicolumn{7}{l}{\textit{(a) RF-CE (MLS)}} \\[1pt]
\textendash{} & \textendash{} & \sjNA & 80.88 & 78.81 & 95.57 & 74.39 \\
\textendash{} & $\checkmark$  & \sjNA & 81.22 & 79.64 & 96.07 & 68.12 \\
$\checkmark$  & \textendash{} & \sjNA & 76.07 & 75.62 & \sjSecond{97.71} & 82.48 \\
$\checkmark$  & $\checkmark$  & \sjNA & 76.26 & 75.88 & \sjBest{97.81} & 83.99 \\
\addlinespace[2.5pt]
\multicolumn{7}{l}{\textit{(b) OpenSFT (Class-$z$)}} \\[1pt]
\textendash{} & \textendash{} & 97.99 & 57.51 & 55.74 & 94.11 & 86.26 \\
\textendash{} & $\checkmark$  & 97.76 & \sjBest{92.83} & \sjBest{90.10} & 96.03 & \sjBest{40.64} \\
$\checkmark$  & \textendash{} & 98.55 & 66.35 & 58.49 & 82.40 & 82.48 \\
\rowcolor{sjOursBlue}
$\checkmark$  & $\checkmark$  & 98.34 & \sjSecond{89.53} & \sjSecond{87.91} & 97.46 & \sjSecond{50.27} \\
\bottomrule
\end{tabularx}
\caption{Ablation results on u4 across LM initialization, Stage~I, and downstream interface. \textbf{Sub-U} denotes Selection-Val subset \texttt{UNKNOWN} accuracy; bold and underlined numbers mark the best and second-best test scores.}
\label{tab:pretraining}
\end{table}

To answer RQ2, Table~\ref{tab:pretraining} runs a $2\times 2\times 2$ u4 ablation across LM initialization, Stage~I, and downstream interface. All eight variants train on a single GPU, contrasting OpenSFT against prompt-free \textbf{RF-CE}, which mean-pools the 40 RF tokens into a linear $K$-way head scored by MLS; details appear in Appendix~\ref{app:impl}.

\paragraph{Pretraining converts partial-list shortcuts into unseen-class transfer.}
On half-subset validation, both random and pretrained OpenSFT exceed 97.7\% Sub-U: like a closed-set classifier paired with a set-membership check, a model can fit subset omission without language priors. On full-candidate unseen tests, however, random initialization collapses (57.51--66.35 AUROC), whereas pretraining lifts AUROC by over 23 points and cuts FPR95 by over 32 points. Under signal-only RF-CE, the same pretrained weights shift AUROC by under 0.5 points, showing that language pretraining prevents partial-list shortcuts specifically through the candidate prompt and frozen native vocabulary head.

\paragraph{Stage~I tightens known clusters without automatically widening open-set margins.}
On pretrained weights, Stage~I consistently boosts known accuracy in both RF-CE (+1.74) and OpenSFT (+1.43), confirming that masked waveform reconstruction sharpens familiar-class representations. Yet omitting Stage~I yields higher u4 AUROC in both paths (+4.96 and +3.30), while on random OpenSFT Stage~I trades closed-set accuracy for modest unseen separation. Label-free reconstruction thus acts primarily as a closed-set regularizer, motivating open-set-aware pretraining in future work.

\section{Analysis}
\label{sec:analysis}

\subsection{Score Calibration and Margin Geometry}
\label{sec:calibration_analysis}
\begin{table}[t]
\centering\scriptsize
\setlength{\tabcolsep}{2.2pt}
\renewcommand{\arraystretch}{1.15}
\begin{tabularx}{\columnwidth}{l*{6}{>{\centering\arraybackslash}X}}
\toprule
\rowcolor{sjHeaderGray}
& \multicolumn{2}{c}{AUROC ($\uparrow$)} & \multicolumn{2}{c}{OSCR ($\uparrow$)} & \multicolumn{2}{c}{FPR95 ($\downarrow$)} \\
\cmidrule(lr){2-3}\cmidrule(lr){4-5}\cmidrule(lr){6-7}
\rowcolor{sjHeaderGray}
Protocol & Raw & Class-$z$ & Raw & Class-$z$ & Raw & Class-$z$ \\
\midrule
u2 & 88.82 & \sjBest{90.54} & 87.25 & \sjBest{88.66} & \sjBest{48.60} & 55.51 \\
u3 & 69.24 & \sjBest{74.27} & 69.00 & \sjBest{73.49} & 87.52 & \sjBest{70.89} \\
u4 & 78.46 & \sjBest{91.61} & 77.61 & \sjBest{90.23} & 59.89 & \sjBest{41.51} \\
u5 & 72.82 & \sjBest{81.95} & 72.04 & \sjBest{81.18} & 99.48 & \sjBest{89.97} \\
\midrule
\rowcolor{sjOursBlue}
\textbf{Mean} & 77.33 & \sjBest{84.59} & 76.48 & \sjBest{83.39} & 73.87 & \sjBest{64.47} \\
\rowcolor{sjOursBlue}
\textbf{$\Delta$} & \sjNA & \sjBest{+7.26} & \sjNA & \sjBest{+6.91} & \sjNA & \sjBest{$-$9.40} \\
\bottomrule
\end{tabularx}
\caption{Comparison of raw margins $m(x)$ and predicted-class $z$-standardization across u2--u5. Bold numbers indicate the better score in each pair.}
\label{tab:calibration_transfer}
\end{table}

\begin{figure}[t]
    \centering
    \includegraphics[width=\columnwidth]{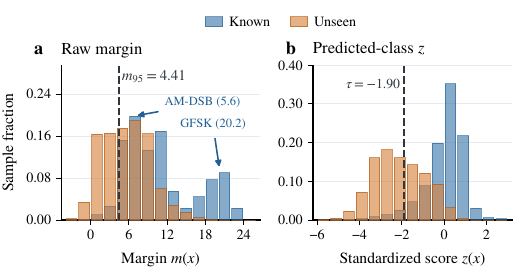}
    \caption{Known versus unseen test score distributions on u4 for (a)~raw margins $m(x)$ with the 95\%-TPR operating point $m_{95}$ and (b)~predicted-class $z$-scores ($>99.6\%$ within $[-6,3]$) with Known-Cal threshold $\tau$. Standardization aligns known modes near zero and separates unseen waveforms below $\tau$.}
    \label{fig:score_dist}
\end{figure}

Table~\ref{tab:calibration_transfer} isolates Eq.~\eqref{eq:class_z} across all four protocols: predicted-class $z$-standardization improves AUROC and OSCR on every split (mean +7.26 AUROC, +6.91 OSCR) and cuts FPR95 substantially on the three harder protocols u3--u5 (mean $-$9.40 across u2--u5). Even before standardization, OSCAL's raw u4 margin (78.46 AUROC) already surpasses Prompted-CE MLS (59.56) by +18.90 points, confirming that OpenSFT improves open-set separation prior to calibration. Figure~\ref{fig:score_dist} illustrates the underlying geometry on representative protocol u4 (all remaining protocols and Known-Cal scale ranges in Appendix~\ref{app:score_dists}): in raw margin space (Figure~\ref{fig:score_dist}a), known waveforms form a broad multimodal distribution because analog schemes produce much smaller margins (AM-DSB averages 5.6) than distinct digital schemes (GFSK averages 20.2), burying unseen signals inside the lower known mode. Standardizing by predicted-class Known-Cal statistics (Figure~\ref{fig:score_dist}b) aligns known classes near zero and shifts unseen scores sharply negative relative to $\tau=-1.90$.

\subsection{Physical Proximity and Operating-Point Dissection}
\label{sec:error_analysis}
\begin{table}[t]
\centering\scriptsize
\setlength{\tabcolsep}{2.6pt}
\renewcommand{\arraystretch}{1.09}
\begin{tabularx}{\columnwidth}{cc>{\centering\arraybackslash}Xcc}
\toprule
\rowcolor{sjHeaderGray}
Protocol & \shortstack{Unseen\\Class} & \shortstack{Dominant Known\\Absorber (\%)} & \shortstack{AUC\\$\uparrow$} & \shortstack{Mean $z$\\$\downarrow$} \\
\midrule
\multirow{2}{*}{u2}
& QAM16  & 8PSK (57.4), PAM4 (25.7) & 90.02 & $-1.92$ \\
& QAM64  & 8PSK (51.5), PAM4 (31.6) & 91.06 & $-2.04$ \\
\midrule
\multirow{3}{*}{u3}
& 8PSK   & QPSK (98.8)              & 46.99 & $+0.07$ \\
& QAM16  & QPSK (93.3)              & 86.40 & $-1.66$ \\
& QAM64  & QPSK (90.9)              & 89.40 & $-1.97$ \\
\midrule
\multirow{4}{*}{u4}
& 8PSK   & BPSK (86.7)              & 89.63 & $-1.96$ \\
& QAM16  & PAM4 (83.9)              & 95.18 & $-2.49$ \\
& QAM64  & PAM4 (87.2)              & 94.23 & $-2.32$ \\
& QPSK   & BPSK (87.1)              & 87.41 & $-1.80$ \\
\midrule
\multirow{5}{*}{u5}
& AM-DSB & WBFM (99.9)              & 82.36 & $-0.97$ \\
& 8PSK   & BPSK (95.9)              & 84.25 & $-1.00$ \\
& QPSK   & BPSK (94.1)              & 83.11 & $-0.95$ \\
& QAM16  & CPFSK (61.6)             & 80.11 & $-1.29$ \\
& QAM64  & CPFSK (65.8)             & 79.89 & $-1.28$ \\
\bottomrule
\end{tabularx}
\caption{Per-class unseen absorption and threshold-free separation across u2--u5. \textit{AUC} and \textit{Mean $z$} evaluate each unseen class's standardized score against all known test waveforms (known mean $z \approx 0$).}
\label{tab:unseen_absorption}
\end{table}

\paragraph{Physical proximity governs unseen-class separation.}
Table~\ref{tab:unseen_absorption} dissects every held-out modulation across u2--u5 along two threshold-free axes: which known class absorbs it prior to thresholding ($\hat{y}(x)$) and how cleanly its standardized score separates from known waveforms (per-class AUC and mean $z$). Across splits, routing absorption follows modulation physics, yet only a true near-duplicate collapses score ranking. On u3, 98.8\% of unseen 8PSK waveforms are absorbed by their closest phase-modulation neighbor QPSK, collapsing per-class AUC to 46.99 (mean $z=+0.07$) and explaining why u3 is the hardest split across methods. Once QPSK is also held out in u4, 8PSK separation rebounds immediately to 89.63 AUC (mean $z=-1.96$), alongside 94.23--95.18 AUC on QAM16/QAM64. Crucially, on u5, even though 99.9\% of unseen analog AM-DSB waveforms are routed to WBFM before thresholding, all five unseen classes maintain consistent separation (79.89--84.25 AUC; mean $z \in [-1.29, -0.95]$), confirming intact open-set ranking across every held-out modulation.

\paragraph{Ranking versus operating-point decoupling on u5.}
Table~\ref{tab:unseen_absorption} thus explains why OSCAL leads u5 in AUROC (81.95) despite elevated FPR95 (89.97): the gap lies at the single 95\%-TPR threshold rather than unseen ranking. Under u5's compact 3-vs.-3 subsets, subset validation saturates while full-candidate PAM4 accuracy fluctuates across epochs, as detailed in Appendix~\ref{app:u5_diag}. Selecting strictly by subset SFT loss picks an early checkpoint during a transient PAM4 dip, misrouting over one-third of known PAM4 waveforms into higher-mean digital buckets such as GFSK, CPFSK, and BPSK with deeply negative raw margins. Standardizing against those high bucket means creates a negative known $z(x)$ tail that pulls $\tau$ to $-2.21$ and inflates $\sigma_k$, compressing unseen $z$-scores into $[-1.29, -0.95]$ above $\tau$ (Figure~\ref{fig:score_dist_other}f).

\subsection{Discussion and Implications}
\label{sec:discussion}
Our findings highlight two lessons for non-text open-world LM adaptation. First, language pretraining contributes neither as a generic signal encoder nor by fitting partial lists, but by enabling \emph{out-of-context abstention transfer} through the frozen native vocabulary head. Second, reliable operating points require co-designing representation, predicted-class calibration, and half-to-full candidate selection.

More broadly, extending abstention learning such as SEAL \citep{huang2025seal} to continuous waveforms bridges signal recognition with multiple-choice symbol biases \citep{zheng2024mcqa} and ``None of the Above'' failure modes \citep{tam2025nota}. Grounding abstention in native vocabulary anchors avoids the split-specific collapses of detached OSR modules. At the same time, our prompt-intervention study in Appendix~\ref{app:semantic} shows that standard OpenSFT binds waveforms mainly to fixed code--name anchors rather than zero-shot modulation descriptions, whereas partially anonymizing class names during training enables genuine description reading at the cost of narrower open-set margins. Reconciling symbolic anchor stability with lexical-free semantic grounding thus remains an open direction.

\section{Conclusion}
We presented OSCAL, an abstention learning framework that initializes continuous dual-view RF tokens via masked reconstruction and teaches a frozen pretrained LM to reject unseen modulations through complementary candidate-conditioned instruction tuning and predicted-class $z$-standardization. Across four RML22 protocols and 18 baselines, OSCAL achieves the strongest cross-protocol robustness without split-specific collapses. Controlled ablations show that language pretraining turns partial-list abstention into genuine unseen-class transfer, while our analyses characterize the calibration, physical, and contextual boundaries of non-text abstention.

\section*{Limitations}
This work exhibits several methodological limitations worth noting. Firstly, Stage~II constructs complementary candidate lists via uniform random half-partitions with fixed alphabetic codes and static modulation descriptions. While simple and effective, this design creates a candidate-length gap between half-subset training/selection and full-candidate deployment (as diagnosed on u5) and encourages reliance on fixed symbolic code--name anchors rather than flexible description reading (Appendix~\ref{app:semantic}); exploring variable subset sizes, full-candidate auxiliary selection, and lexical-free semantic grounding represents an important direction for future work. Secondly, Stage~I masked reconstruction and Stage~II OpenSFT are optimized sequentially without an explicit open-set representation objective; as shown in the u4 ablation, pure reconstruction sharpens closed-set accuracy without necessarily widening open-set margins, pointing toward joint or open-set-aware pretraining objectives. Thirdly, inference relies on post-hoc predicted-class $z$-standardization with a single global Known-Cal quantile threshold, and processing RF tokens jointly with full candidate prompts introduces higher memory and forward-pass cost than compact CNN classifiers. Exploring end-to-end calibrated abstention boundaries, adaptive per-class thresholds, and prompt-prefix caching or distillation offers promising avenues for future improvement.

\bibliography{custom}

\clearpage
\appendix

\section{Architecture, Training, and Calibration Details}
\label{app:impl}

\paragraph{Dual-view RF frontend and masking.}
Each complex IQ waveform $x \in \mathbb{R}^{128 \times 2}$ is normalized by its complex RMS ($\epsilon = 10^{-8}$) without DC removal. The temporal branch applies a stride-4 \texttt{Conv1d} ($k=4$, width $d=256$) to map the 128 IQ samples into 32 temporal tokens. In parallel, the spectral branch computes a 32-point Hann-window STFT (window length 32, hop length 8, \texttt{center=false}, two-sided, frequency-shifted via \texttt{fftshift}), averages squared magnitude across time, applies $\log(1+\cdot)$, and passes the resulting 32 frequency bins through a second stride-4 \texttt{Conv1d} ($k=4$, $d=256$) to yield 8 PSD tokens, forming a 40-token RF prefix. Each token receives learned view embeddings, learned token-position embeddings, and a bias-free linear projection of normalized patch centers $(t_i, 0)$ for IQ and $(0, f_j)$ for PSD in $[-1, 1]^2$.

During Stage~I, we mask 12 IQ tokens (contiguous spans of length 2--4; ratio 0.40) and 2 PSD tokens (a contiguous band of length 2--3; ratio 0.30) using learned view-specific mask embeddings. A linear input bridge ($256 \to 1024$) projects the 40 RF tokens into frozen \texttt{Qwen3-0.6B-Base} (28 layers, hidden size 1024) under full bidirectional attention. A Stage~I-only linear output bridge ($1024 \to 256$) and view-specific linear heads reconstruct masked IQ ($d_{\mathrm{IQ}}=8$) and PSD ($d_{\mathrm{PSD}}=4$) patches. Targets $t_i^v$ are standardized per sample and view by subtracting the mean and dividing by $\sqrt{s_v^2 + 10^{-4}}$ (stop-gradient), where $s_v^2$ is the sample--view population variance.

\paragraph{LoRA, optimization, and selection.}
LoRA adapters ($r=16$, $\alpha=32$, dropout 0.05) are attached to all seven attention and feed-forward projections (\texttt{q\_proj, k\_proj, v\_proj, o\_proj, gate\_proj, up\_proj, down\_proj}) in every layer, totaling 10.4M trainable parameters (1.7\% of 606.7M) together with the RF frontend and input bridge; all base LM weights and tied input/output embeddings stay frozen. In Stage~II, candidate answers are scored in FP32 over the full Qwen3 vocabulary without appending an EOS token, and every alphabetic code (\texttt{A}--\texttt{K}, \texttt{U}) is verified to tokenize as a distinct single token at the \texttt{Answer:} boundary without prompt-token merging. Full optimization hyperparameters are summarized in Table~\ref{tab:hyperparams}. All runs use training seed 42 and fixed-batch SDPA scoring with batch size 16, protocol-fixed padding, and BF16 precision. Stage~II applies early stopping with a 3-epoch patience on Selection-Val SFT loss, selecting the minimum-loss checkpoint and breaking ties by full-candidate known accuracy and earlier epoch. Main-table OSCAL runs train with 8-GPU DDP, whereas all eight variants in Table~\ref{tab:pretraining} train Stage~I and downstream stages end-to-end on a single GPU under the same early-stopping rule, accounting for the minor u4 difference between Tables~\ref{tab:openset_main} and~\ref{tab:pretraining}. For RF-CE, we mean-pool all 40 RF tokens into a linear $K$-way head with learning rate $10^{-3}$.

\paragraph{Calibration fallbacks.}
In Eq.~\eqref{eq:class_z}, if a predicted-class Known-Cal bucket contains fewer than 50 samples or has population standard deviation $\sigma_k < 10^{-6}$, OSCAL falls back to the global Known-Cal mean and standard deviation. The acceptance threshold $\tau$ is computed as the 5th percentile of Known-Cal $z(x)$ scores via linear quantile interpolation.

\begin{table}[t]
\centering\scriptsize
\setlength{\tabcolsep}{3.5pt}
\renewcommand{\arraystretch}{1.15}
\begin{tabularx}{\columnwidth}{l>{\raggedright\arraybackslash}X>{\raggedright\arraybackslash}X}
\toprule
\rowcolor{sjHeaderGray}
Hyperparameter & Stage~I (Recon.) & Stage~II (OpenSFT) \\
\midrule
Epochs (max) & 50 & 30 \\
Early stopping & None & 3-epoch patience \\
Effective RF batch & 256 & 64 (2 tasks / RF) \\
Optimizer & AdamW, $(0.9, 0.95)$ & AdamW, $(0.9, 0.999)$ \\
Learning rate & $3\times 10^{-4}$ & RF/bridge $3\times 10^{-5}$; LoRA $10^{-4}$ \\
Weight decay & 0.05 & 0.0 \\
Gradient clip & 1.0 & None \\
LR schedule & 5\% warmup, cosine (min $10^{-5}$) & 3\% warmup, cosine (min $0.1\times$) \\
Precision & BF16 & BF16 (FP32 vocab head) \\
LoRA config & \multicolumn{2}{l}{$r{=}16$, $\alpha{=}32$, dropout $0.05$ ($7{\times}28$ proj.)} \\
Selection rule & Min Val $\mathcal{L}_{\mathrm{rec}}$ & Min Val $\mathcal{L}_{\mathrm{SFT}}$; ties: Acc, epoch \\
\bottomrule
\end{tabularx}
\caption{Optimization and adaptation hyperparameters for Stage~I and Stage~II of OSCAL.}
\label{tab:hyperparams}
\end{table}

\section{Prompt Templates, Modulation Descriptions, and Protocol Splits}
\label{app:prompts}

\paragraph{Exact prompt template.}
Every Stage~II training pair and full-candidate inference query uses the exact instruction template below, prefixed by the 40 continuous RF tokens:
\begin{quote}\scriptsize\ttfamily
You are an expert in radio modulation classification.\\
Analyze the input RF signal and select its modulation type from the given candidate categories.\\
Candidate categories:\\
\{candidates\}\\
Which category does the RF signal belong to? Answer:
\end{quote}
Each line in \texttt{\{candidates\}} takes the form \texttt{<Code>. <Description>}, followed by \texttt{U. UNKNOWN: None of the listed modulation types.} in open-set prompts. Table~\ref{tab:descriptions} lists all 11 invariant global class codes and verbatim physical modulation descriptions used across protocols.

\begin{table*}[t]
\centering\scriptsize
\setlength{\tabcolsep}{4pt}
\renewcommand{\arraystretch}{1.15}
\begin{tabularx}{\textwidth}{clX}
\toprule
\rowcolor{sjHeaderGray}
Code & Class & Verbatim Modulation Description in Prompt \\
\midrule
\texttt{A} & 8PSK & \texttt{8PSK: digital phase-shift keying (PSK family); 8 equally spaced carrier phase states; constant envelope; double-sideband spectrum.} \\
\texttt{B} & AM-DSB & \texttt{AM-DSB: analog amplitude modulation, double sideband with carrier; information carried by the envelope amplitude; non-constant envelope; spectrum symmetric around the carrier.} \\
\texttt{C} & AM-SSB & \texttt{AM-SSB: analog amplitude modulation, single sideband; information carried by the envelope amplitude; non-constant envelope; spectrum on one side of the carrier only.} \\
\texttt{D} & BPSK & \texttt{BPSK: digital phase-shift keying (PSK family); 2 carrier phase states; constant envelope; double-sideband spectrum.} \\
\texttt{E} & CPFSK & \texttt{CPFSK: digital frequency-shift keying (FSK family); 2 frequency tones with continuous phase at symbol boundaries; constant envelope; energy concentrated around two tone frequencies.} \\
\texttt{F} & GFSK & \texttt{GFSK: digital frequency-shift keying (FSK family) with Gaussian pulse shaping; 2 frequency tones with smooth frequency transitions; constant envelope; compact spectrum.} \\
\texttt{G} & PAM4 & \texttt{PAM4: digital pulse amplitude modulation; 4 real amplitude levels on the in-phase axis only; non-constant envelope; double-sideband spectrum.} \\
\texttt{H} & QAM16 & \texttt{QAM16: digital quadrature amplitude modulation (QAM family); 16-point square constellation with 3 amplitude rings; non-constant envelope; double-sideband spectrum.} \\
\texttt{I} & QAM64 & \texttt{QAM64: digital quadrature amplitude modulation (QAM family); 64-point square constellation with 9 amplitude rings; non-constant envelope; double-sideband spectrum.} \\
\texttt{J} & QPSK & \texttt{QPSK: digital phase-shift keying (PSK family); 4 equally spaced carrier phase states; constant envelope; double-sideband spectrum.} \\
\texttt{K} & WBFM & \texttt{WBFM: analog wideband frequency modulation; information carried by the instantaneous frequency; constant envelope; wide spectrum.} \\
\texttt{U} & UNKNOWN & \texttt{UNKNOWN: None of the listed modulation types.} \\
\bottomrule
\end{tabularx}
\caption{Global single-token answer codes and verbatim physical modulation descriptions for all 11 RML22 classes and \texttt{UNKNOWN}.}
\label{tab:descriptions}
\end{table*}

\paragraph{Protocol data splits.}
The 0--20~dB subset of RML22 (2~dB steps, 11 SNRs) is partitioned once using a fixed seed-42 stratified 60\%/20\%/20\% split. Each known class provides 13,200 training samples (1,200 per SNR), 2,200 Selection-Val samples, and 2,200 Known-Cal samples. Every test fold contains all 11 classes (48,400 samples; 400 per class--SNR). Table~\ref{tab:protocol_stats} summarizes the nested u2--u5 partitions.

\begin{table*}[t]
\centering\scriptsize
\setlength{\tabcolsep}{4.5pt}
\renewcommand{\arraystretch}{1.15}
\begin{tabularx}{\textwidth}{cccXrrrr}
\toprule
\rowcolor{sjHeaderGray}
Protocol & $|\mathcal{K}|$ & Half Split & Unseen Classes $\mathcal{U}$ & Train & Sel-Val & Known-Cal & Test (Known / Unseen) \\
\midrule
u2 & 9 & $4+5$ & QAM16, QAM64 & 118,800 & 19,800 & 19,800 & 48,400 (39,600 / 8,800) \\
u3 & 8 & $4+4$ & 8PSK, QAM16, QAM64 & 105,600 & 17,600 & 17,600 & 48,400 (35,200 / 13,200) \\
u4 & 7 & $3+4$ & 8PSK, QAM16, QAM64, QPSK & 92,400 & 15,400 & 15,400 & 48,400 (30,800 / 17,600) \\
u5 & 6 & $3+3$ & 8PSK, AM-DSB, QAM16, QAM64, QPSK & 79,200 & 13,200 & 13,200 & 48,400 (26,400 / 22,000) \\
\bottomrule
\end{tabularx}
\caption{Summary of the four nested RML22 open-set protocols (0--20~dB, seed 42). Unseen classes appear strictly in the Test fold.}
\label{tab:protocol_stats}
\end{table*}

\section{Baseline Configurations}
\label{app:baselines}
All 18 baselines are trained and evaluated on the identical seed-42 splits with zero unseen leakage using their official continuous open-set scores.

\paragraph{Group~(a): Prompted-CE post-hoc scores.}
Starting from the exact same Stage~I checkpoint as OSCAL, we fine-tune with closed-set cross-entropy over the full known prompt $P(\mathcal{K})$ (omitting \texttt{UNKNOWN}) using the frozen tied native vocabulary rows for $\mathcal{K}$, selecting by Selection-Val known accuracy (ties broken by CE loss, then epoch). Post-hoc scores compute MSP, MLS, negative Energy, and Weibull-calibrated OpenMax ($\text{tailsize}=20$, $\alpha=3$) on the $K$ known logits.

\paragraph{Group~(b): Qwen-adapted OSR modules.}
CSSR (latent dim 64, $\gamma=0.1$), MEDAF (gate/logit temperature 100, loss weights $[0.7, 1.0, 0.01]$), and CSGRL (latent dim 64, $\gamma=0.1$, margin 5, score weight 0.5, 6 generation epochs) attach their respective OSR modules to the shared Stage~I RF-Qwen backbone (512D features, 30 epochs, effective batch 64, RF LR $3\times 10^{-5}$, LoRA LR $10^{-4}$, head LR $10^{-3}$).

\paragraph{Group~(c): AMC-specific OSR networks.}
We reproduce each published recipe on identical splits: CFSR (ResNet18-1D, AdamW LR $3\times 10^{-3}$ cosine, 50 epochs, batch 64, $\alpha=0.1$, $\eta=0.95$), LGFSCN (LGFE backbone, SGD LR $10^{-3}$, momentum 0.9, 1,000 epochs, batch 128, $\lambda=0.1$), and MLFDL (MCLDNN backbone, Adam LR $2\times 10^{-3}$, 1,000 epochs, batch 256, $\tau=0.1$, $\lambda_{\mathrm{CL}}=0.1$, $\lambda_{\mathrm{MMFC}}=10^{-3}$, manifold mixup $\lambda \sim \mathcal{U}[0.98, 1)$).

\paragraph{Group~(d): Text-prompting LLM AMC pipelines.}
We evaluate Context Is King with 1 example per class and 5-decimal verbalized signal statistics alongside three DiSC-AMC variants built on the official repository's DINO ViT-B/8 (\texttt{facebook/dino-vitb8}) visual autoencoder and classifier (referred to as DINOv2 in \citealp{disc_amc}), trained strictly on known classes with 5 discretization bins and a top-5 shortlist: \texttt{TF, Centroid}, which reproduces the centroid-shortlist 1-shot setting in Table~VI of \citet{disc_amc}; \texttt{TF, RAG}, which uses a FAISS-filled $k_{\mathrm{NN}}{=}50$ visual shortlist paired with stats-space RAG retrieval ($k_{\mathrm{RAG}}{=}10$, selecting 1 exemplar per shortlisted class); and \texttt{QLoRA, RAG}, which fine-tunes 4-bit NF4 QLoRA adapters ($r{=}16$, $\alpha{=}16$, LR $5\times 10^{-5}$, 5 epochs) on discretized 10-D PCA visual embeddings from class-balanced known subsets. Training-free variants evaluate \texttt{Qwen/Qwen3-0.6B-Base} and \texttt{deepseek-ai/DeepSeek-R1-Distill-Qwen-7B} in BF16 via full-vocabulary class-name sequence log-probability scoring, while \texttt{QLoRA, RAG} scores the learned completion sequence per shortlisted class under the 4-bit NF4 base with BF16 compute.

\section{Supplementary Diagnostic Analyses}
\label{app:supp}

\subsection{Score Distributions and Calibration Scales Across Protocols}
\label{app:score_dists}
\begin{figure*}[t]
    \centering
    \includegraphics[width=0.86\textwidth]{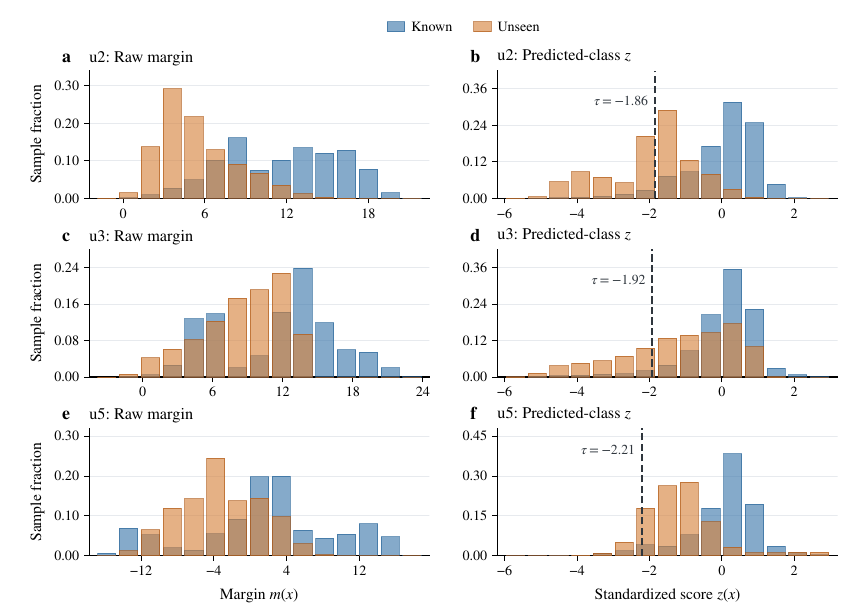}
    \caption{Known versus unseen test score distributions on protocols u2 (a--b), u3 (c--d), and u5 (e--f), comparing raw margins $m(x)$ (left; 100\% coverage) against predicted-class $z$-scores (right; $>99.1\%$ coverage within $[-6,3]$) and the Known-Cal 95\%-TPR threshold $\tau$.}
    \label{fig:score_dist_other}
\end{figure*}

Complementing the representative u4 visualization in Figure~\ref{fig:score_dist}, Figure~\ref{fig:score_dist_other} plots raw-margin versus predicted-class $z$-score distributions for the remaining protocols (u2, u3, and u5), and Table~\ref{tab:cal_ranges} summarizes the Known-Cal raw-margin dispersion across all four splits. Across u2--u4, analog AM classes consistently exhibit the smallest Known-Cal margins (+4.60 to +7.36) while distinct digital schemes (CPFSK/GFSK) exceed +16.50 to +20.31, explaining the broad multimodal known distributions in raw margin space. On u5, PAM4's transient dip ($-11.09$) creates an additional negative known mode in raw space (Figure~\ref{fig:score_dist_other}e); predicted-class $z$-standardization re-centers known classes near zero and cleanly separates the unseen distribution around $z \approx -1.0$ (Figure~\ref{fig:score_dist_other}f), while the shifted threshold $\tau=-2.21$ sits below the unseen cluster.

\begin{table}[t]
\centering\scriptsize
\setlength{\tabcolsep}{3.0pt}
\renewcommand{\arraystretch}{1.14}
\begin{tabularx}{\columnwidth}{c>{\centering\arraybackslash}X>{\centering\arraybackslash}Xcc}
\toprule
\rowcolor{sjHeaderGray}
Protocol & Lowest Bucket Mean & Highest Bucket Mean & Global $\sigma$ & $\tau$ \\
\midrule
u2 & AM-DSB ($+7.36$) & CPFSK ($+16.50$) & 4.17 & $-1.86$ \\
u3 & AM-DSB ($+4.60$) & GFSK ($+18.09$)  & 4.86 & $-1.92$ \\
u4 & AM-DSB ($+5.52$) & GFSK ($+20.31$)  & 5.53 & $-1.90$ \\
u5 & PAM4 ($-11.09$)  & GFSK ($+8.65$)   & 7.65 & $-2.21$ \\
\bottomrule
\end{tabularx}
\caption{Known-Cal predicted-class raw-margin scale dispersion across u2--u5 (lowest and highest class bucket means $\mu_k$, global raw-margin standard deviation, and calibrated 5th-percentile $z$-threshold $\tau$). On u4, the corresponding test-set class means annotated in Figure~\ref{fig:score_dist}a are $+5.60$ (AM-DSB) and $+20.24$ (GFSK).}
\label{tab:cal_ranges}
\end{table}

\subsection{u5 Selection Trajectory and Per-Class Calibration}
\label{app:u5_diag}
Table~\ref{tab:u5_trajectory} documents the u5 Selection-Val trajectory alongside per-class test recalls and Known-Cal raw margins at the selected epoch~1 checkpoint. While half-subset validation saturates across all five epochs at 98.81--99.01\% Sub-ID and 99.17--99.40\% Sub-U, full-candidate known accuracy fluctuates markedly, starting at 97.48\% at epoch~0, dipping to 93.37\% at epoch~1 and 84.10\% at epoch~3, and recovering to 98.33\% at epoch~4 before patience-3 early stopping terminates training. At epoch~1, five of six known classes retain 98.70--99.75\% pre-rejection test recall, whereas PAM4 drops to 62.75\%. Out of 2,200 Known-Cal PAM4 waveforms, only 1,366 enter the PAM4 bucket while 834 negative-margin samples are misrouted into higher-mean buckets, expanding GFSK, CPFSK, and BPSK bucket counts to 2,613, 2,451, and 2,367. Standardizing those misrouted samples against higher bucket means produces the negative $z$-tail that shifts $\tau$ to $-2.21$ and inflates $\sigma_k$ across digital buckets.

\begin{table}[t]
\centering\scriptsize
\setlength{\tabcolsep}{3.5pt}
\renewcommand{\arraystretch}{1.14}
\begin{tabularx}{\columnwidth}{c*{4}{>{\centering\arraybackslash}X}}
\toprule
\rowcolor{sjHeaderGray}
\multicolumn{5}{l}{\textit{(a) u5 Selection-Val trajectory across epochs}} \\
\midrule
Epoch & SFT Loss ($\downarrow$) & Sub-ID & Sub-U & Full Acc \\
\midrule
0 & 0.0390 & 98.88 & 99.17 & 97.48 \\
\rowcolor{sjOursBlue}
\textbf{1*} & \textbf{0.0269} & \textbf{99.01} & \textbf{99.40} & \textbf{93.37} \\
2 & 0.0314 & 98.81 & 99.29 & 91.51 \\
3 & 0.0344 & 98.81 & 99.27 & 84.10 \\
4 & 0.0308 & 98.98 & 99.36 & 98.33 \\
\midrule
\rowcolor{sjHeaderGray}
\multicolumn{5}{l}{\textit{(b) Per-class recall and Known-Cal buckets at epoch~1}} \\
\midrule
Class & Recall (\%) & Cal $n$ & Mean $m(x)$ & Std $m(x)$ \\
\midrule
AM-SSB & 99.75 & 2,221 & $+2.23$ & 0.94 \\
BPSK   & 99.64 & 2,367 & $+5.48$ & 4.41 \\
CPFSK  & 99.20 & 2,451 & $-0.09$ & 4.73 \\
GFSK   & 99.43 & 2,613 & $+8.65$ & 9.56 \\
\textbf{PAM4} & \textbf{62.75} & \textbf{1,366} & $\mathbf{-11.09}$ & \textbf{2.37} \\
WBFM   & 98.70 & 2,182 & $-1.47$ & 2.09 \\
\bottomrule
\end{tabularx}
\caption{Diagnostic breakdown of u5 half-to-full selection mismatch, where * marks the epoch selected by minimum Selection-Val SFT loss and \textbf{Sub-ID}/\textbf{Sub-U} denote half-subset known and \texttt{UNKNOWN} validation accuracy. In part~(b), \textit{Recall} reports true-class pre-rejection test accuracy, whereas \textit{Cal $n$}, \textit{Mean $m(x)$}, and \textit{Std $m(x)$} report Known-Cal predicted-class bucket statistics.}
\label{tab:u5_trajectory}
\end{table}

\begin{table*}[tb]
\centering\footnotesize
\setlength{\tabcolsep}{4.0pt}
\renewcommand{\arraystretch}{1.16}
\begin{tabularx}{\textwidth}{l*{10}{>{\centering\arraybackslash}X}}
\toprule
\rowcolor{sjHeaderGray}
& \multicolumn{3}{c}{Original} & \multicolumn{3}{c}{Paraphrase (avg.\ 2)} & \multicolumn{2}{c}{Swapped (avg.\ 3)} & \multicolumn{2}{c}{Anon.\ (avg.\ 2)} \\
\cmidrule(lr){2-4}\cmidrule(lr){5-7}\cmidrule(lr){8-9}\cmidrule(lr){10-11}
\rowcolor{sjHeaderGray}
Model (u5) & Acc & AUC & FPR & Acc & AUC & FPR & Name & Desc & Desc & AUC \\
\midrule
OSCAL (standard) & 93.25 & 81.95 & 89.97 & 99.00 & 87.83 & 54.65 & 97.02 & 1.12 & 0.11 & 90.88 \\
Paraphrase-Aug & 99.03 & 80.57 & 69.19 & 98.87 & 76.19 & 71.62 & 98.92 & 0.28 & 0.22 & 79.86 \\
Paraphrase + DescMix (25\%) & 98.92 & 73.86 & 77.23 & 98.90 & 68.40 & 68.41 & 20.68 & 77.28 & 90.55 & 30.26 \\
\bottomrule
\end{tabularx}
\caption{Prompt-intervention and semantic-grounding probe on u5 under condition-specific Known-Cal calibration. Under Swapped descriptions, \textit{Name} and \textit{Desc} report accuracy aligned to class names versus physical descriptions; under Anonymized labels with class names removed, \textit{Desc} and \textit{AUC} report description-aligned accuracy and open-set AUROC.}
\label{tab:semantic_probe}
\end{table*}

\subsection{Prompt Intervention and Semantic Grounding Probe on u5}
\label{app:semantic}
To test whether OSCAL relies on physical modulation descriptions or fixed code--name bindings, we conduct a controlled prompt-intervention study on u5 comparing three Stage~II models trained from the same Stage~I checkpoint: (i)~standard \textbf{OSCAL} (static descriptions), (ii)~\textbf{Paraphrase-Aug} (sampling from three phrasing variants per class during Stage~II), and (iii)~\textbf{Paraphrase + DescMix (25\%)} (additionally anonymizing class names in 25\% of Stage~II prompts so the model must read descriptions). Each model is evaluated under condition-specific Known-Cal calibration across four prompt suites: \textbf{Original}, \textbf{Paraphrase} (mean of two unseen dev paraphrases), \textbf{Swapped} (mean of three permutations swapping descriptions across class names, measuring accuracy aligned to class names vs.\ descriptions), and \textbf{Anonymized} (mean of two runs stripping class names so only physical descriptions remain).

Table~\ref{tab:semantic_probe} reveals two complementary findings. First, merely paraphrasing descriptions at test time immediately recovers standard OSCAL's u5 operating point (Acc $93.25 \to 99.00\%$, AUROC $81.95 \to 87.83$, FPR95 $89.97 \to 54.65$), independently corroborating that u5's dip stems from brittle surface-context selection rather than degraded representation quality. Second, standard OSCAL and Paraphrase-Aug rely overwhelmingly on fixed code--name bindings (97.02--98.92\% name-aligned accuracy under swapped descriptions; 0.11--0.22\% under anonymized labels). Adding 25\% anonymized description-mixed supervision successfully shifts the model toward genuine description grounding---raising anonymized description accuracy from 0.11\% to 90.55\% and swapped description-aligned accuracy to 77.28\%---yet reduces open-set AUROC on original prompts (81.95 $\to$ 73.86) and degrades anonymized AUROC (30.26). Reconciling lexical-free semantic grounding with strong open-set rejection margins thus remains an open research problem.

\end{document}